\documentclass[letterpaper,titlepage,11pt]{article}
\pdfoutput=1

\usepackage{cite}
\usepackage{amsmath,amssymb,amsthm,mathrsfs,bbm}
\usepackage{latexsym,amscd,amsbsy,amsfonts,dsfont}
\usepackage{graphicx}
\usepackage{float}
\usepackage{ulem}
\usepackage[
      colorlinks=true,
      linkcolor=blue,
      urlcolor=blue,
      filecolor=black,
      citecolor=red,
      pdfstartview=FitV,
      pdftitle={},
        pdfauthor={David Licht, Raimon Luna, Ryotaku Suzuki},
        pdfsubject={},
        pdfkeywords={},
        pdfpagemode=None,
        bookmarksopen=true
      ]{hyperref}\usepackage{caption}

\def\clock{{\count0=\time
           \divide\count0 60
           \ifnum\count0<10 0\fi\the\count0
           \multiply\count0 -60 \advance\count0 \time
           :\ifnum\count0<10 0\fi \the\count0
         }}
\newcommand{\timestamp}{{\small\vbox{\hbox{\tt\jobname.tex}
\hbox{\the\day/\the\month/\the\year, \clock}}}}

\newcommand{\RR}{\mathcal{R}}
\newcommand{\cR}{\mathcal{R}} 

\newcommand{\cS}{{\mathcal S}}

\newcommand{\cM}{{\mathcal M}}

\newcommand{\veps}{\varepsilon}
\newcommand{\ord}[1]{{\mathcal O}\left(#1\right)}
\newcommand{\fr}[1]{\frac{1}{#1}}
\newcommand{\nonum}{\nonumber\\ }

\newcommand{\cout}[1]{}

\newcommand{\ie}{{\it i.e.,\,}}

\newcommand{\bfk}{\boldsymbol{k}}
\newcommand{\bfnu}{\boldsymbol{\nu}}

\usepackage{bm}
\usepackage{latexsym}
\usepackage[latin1]{inputenc}
\usepackage{amsfonts}
\usepackage{amssymb}
\usepackage{amsmath}
\usepackage{subfigure}

\numberwithin{equation}{section}

\begin{document}

\begin{titlepage}
\rightline{TTI-MATHPHYS-9}
\vglue 2cm 

\centerline{\LARGE \bf Lattice Black Branes at Large $D$}

\vskip 1.6cm
\centerline{\bf David Licht$^{a,b}$, Raimon Luna$^{a,c,d}$ and Ryotaku Suzuki$^{a,e,f}$}
\vskip 0.5cm

\centerline{\sl $^{a}$Departament de F{\'\i}sica Qu\`antica i Astrof\'{\i}sica, Institut de Ci\`encies del Cosmos,}
\centerline{\sl  Universitat de Barcelona, Mart\'{\i} i Franqu\`es 1, E-08028 Barcelona, Spain}

\centerline{\sl $^{b}$Department of Physics, Ben-Gurion University of the Negev, Beer-Sheva 84105, Israel}

\centerline{\sl $^{c}$CENTRA, Departamento de F\'{\i}sica, Instituto Superior T\'ecnico - IST,}
\centerline{\sl Universidade de Lisboa - UL, Avenida Rovisco Pais 1, 1049 Lisboa, Portugal}

\centerline{\sl $^{d}$Departamento de Astronom\'{\i}a y Astrof\'{\i}sica, Universitat de Val\`encia,}
\centerline{\sl Dr. Moliner 50, 46100, Burjassot (Val\`encia), Spain}

\centerline{\sl $^{e}$Department of Physics, Osaka City University}
\centerline{\sl Sugimoto 3-3-138, Osaka 558-8585, Japan}

\centerline{\sl $^{f}$Mathematical Physics Laboratory, Toyota Technological Institute}
\centerline{\sl Hisakata 2-12-1, Nagoya 468-8511, Japan}

\vskip 1.cm
\centerline{\bf Abstract} \vskip 0.2cm \noindent

\noindent 
We explore the phase space of non-uniform black branes compactified on oblique lattices with a large number of dimensions. We find the phase diagrams for different periodicities and angles, and determine the thermodynamically preferred phases for each lattice configuration. In a range of angles, we observe that some phases become metastable.

\end{titlepage}
\pagestyle{empty}
\small

\addtocontents{toc}{\protect\setcounter{tocdepth}{2}}
{
	\hypersetup{linkcolor=black,linktoc=all}
	\tableofcontents
}
\normalsize
\newpage
\pagestyle{plain}
\setcounter{page}{1}

\section{Introduction}
\label{sec:introduction}

Periodic deformations of black strings and black branes provide a natural playground to explore the rich phenomena of black holes in higher dimensions. First identified in \cite{Gregory:1993vy, Gregory:1994bj}, the Gregory-Laflamme (GL) instability introduced the possibility of spontaneous breaking of the translational symmetry and opened the door for  cosmic censorship violations (CCV). Evidence of the existence of static non-uniform black strings compactified on a circle was found in \cite{Gubser:2001ac, Wiseman:2002zc}, but such solutions were not adequate as endpoints of the GL instability as they had a lower entropy than the uniform phase \cite{Kudoh:2004hs}. The breakup and transition to localized black holes remained as the only possible endpoint at low dimensions, thus implying a topology change and a CCV. The numerical evolution of the $D=5$ black string turned out to be a continual self-similar cascading into smaller and smaller satellite black holes, which could lead to CCV~\cite{Lehner:2010pn}. The same self-similar cascades leading to the formation of singularities have also been observed in more general setups in $D=5,6,7$~\cite{Figueras:2015hkb,Figueras:2017zwa,Bantilan:2019bvf}.

However, the dynamics in much higher dimensions is different, as it is known that the static non-uniform black string is stabilized above a critical dimension $D_*= 13.5$~\cite{Sorkin:2004qq}. The phase diagrams below and above the critical dimension were carefully studied in \cite{Figueras:2012xj}, revealing the existence of stable non-uniform  black strings above, and in some cases even below, the critical dimension.

Given the rich variety of physical phenomena that appear on the $S^1$ compactification of strings, it is a natural extension of the research to explore lattice deformations of $p$-dimensional black branes compactified on $T^p$. Such analysis was performed in detail by \cite{Dias:2017coo} for 2-branes in $D=6$. Particular attention was placed on the equiangular case, where the torus contains two equilateral triangles. In this particular case, two different arrangements arise from opposite sign excitations of the same zero mode at the GL branching point: triangular an hexagonal lattices. See also \cite{Donos:2013cka, Donos:2015eew} for a similar analysis in asymptotically Anti-de Sitter (AdS) spacetimes.

\begin{figure}[t]
	\begin{center}
		\includegraphics[width=1.1\textwidth]{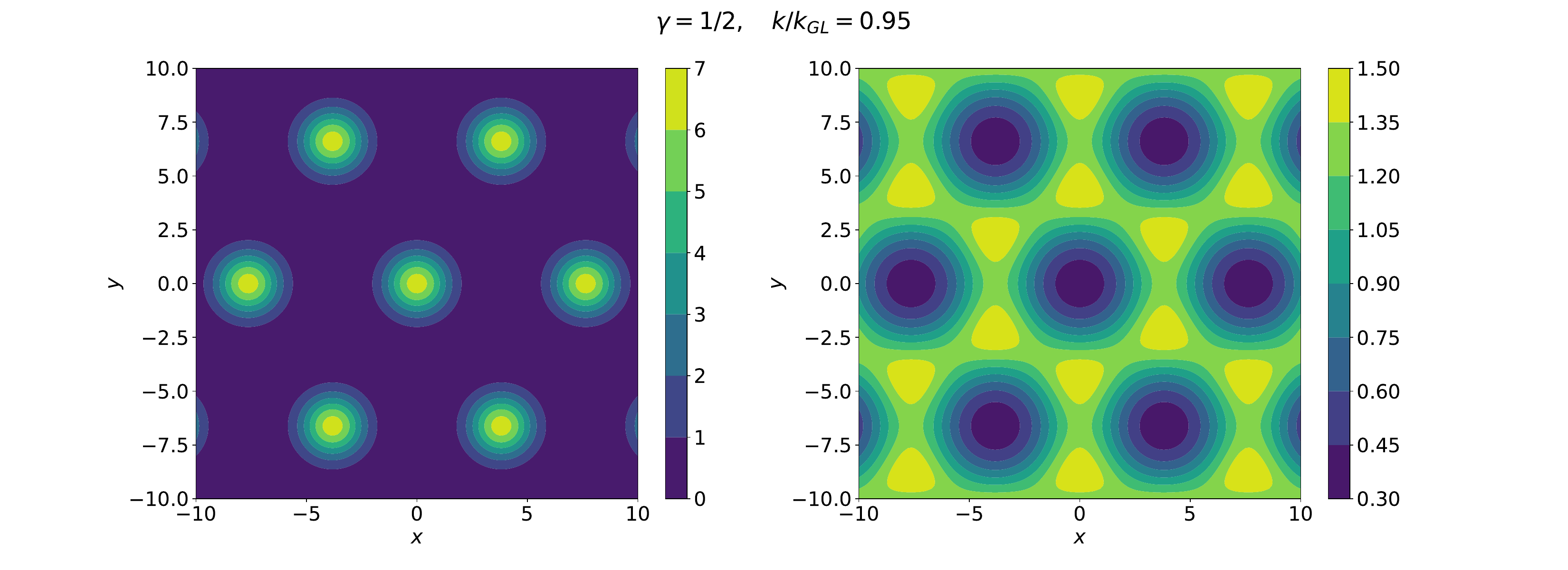}
		\caption{Triangular (left) and hexagonal (right) equiangular lattices at $k = 0.95\, k_{GL}$. The heatmap reperesents the mass density on the black brane.\label{fig:Equiangular_heatmap} }
	\end{center}
\end{figure}

The large $D$ limit is a useful approximation with a wide applicability to grasp analytical features of higher dimensional black holes~\cite{Emparan:2020inr,Asnin:2007rw,Emparan:2013moa}. At large $D$, the gravity of black holes/branes is localized around the thin near-horizon region, which defines a simple effective theory of the horizon deformation dynamics~\cite{Emparan:2015hwa, Suzuki:2015axa,Emparan:2015gva, Emparan:2016sjk,Bhattacharyya:2015dva,Bhattacharyya:2015fdk}. 

Particularly, the large $D$ effective theory approach has made great contributions in understanding the black string dynamics, which include the GL instability and the non-linear evolution to non-uniform phases \cite{Asnin:2007rw,Emparan:2013moa,Emparan:2015hwa,Suzuki:2015axa, Emparan:2016sjk,Emparan:2015gva,Emparan:2018bmi}. The aforementioned critical dimension has also been analytically estimated by using the $1/D$ expansion, with good agreement with the numerical result~\cite{Suzuki:2015axa, Emparan:2018bmi}. While the large $D$ effective theory cannot be applied to the topology-changing transition from the black string phase to the caged black hole phase, this phenomenon can also be studied at large $D$ with a different limit which leads to the Ricci flow equation~\cite{Emparan:2019obu}.

From the success in the black string analysis, it is natural to apply the large $D$ limit to the study of more general black brane instabilities. In ref.~\cite{Rozali:2016yhw}, by solving the large $D$ effective equation with different lattice inclinations, it was shown that the inclination angle plays an important role in the phase of the lattice black brane.

In this paper, we apply the large $D$ effective theory approach to understand the thermodynamics in the rich variety of deformed black $p$-branes compactified in a $p$-dimensional oblique lattice.
We solve the large $D$ effective equation both perturbatively around the branching points from the uniform brane and numerically with $p=2$, with arbitrary wavelengths and angles, and then compare the mass-normalized scale-invariant entropy introduced in~\cite{Andrade:2020ilm} and brane tension to study the thermodynamical properties of lattice solutions. We obtain two important solutions: the {\it hexagonal} and {\it triangular} lattices (see Figure \ref{fig:Equiangular_heatmap}) described in \cite{Dias:2017coo}. We also find another relevant solution, the {\it black stripes}, which presents non-uniformity only in one direction. We show that the black stripes can be thermodynamically stable for a certain range of the angle and periodicity. We also find that either the triangular or the hexagonal phase branches off from the uniform phase, while the other phase branches off from the black stripes. In the equiangular case, both branches merge together in a two-sided branch. The branches of black stripes always start from the uniform phase. We also observe the appearance of a cusp in the triangular phase when the lattice angle $\alpha$ is in the range $\cos\alpha_1=1/\sqrt{10}< \cos\alpha < \cos\alpha_3 \approx 0.57$, in which the stable triangular phase is extended slightly beyond the threshold of the GL instability. More interestingly, other phases shadowed by this extended triangular phase become metastable, \ie thermodynamically not favored but still dynamically stable. This feature was not observed for $D=6$~ \cite{Dias:2017coo}.

The paper is structured as follows: Section \ref{sec:Setup} introduces the large $D$ effective equations and their periodic solutions. In Section \ref{sec:Perturbative analysis} we study perturbatively the lattice configurations for general black $p$-branes distinguishing between equiangular and non-equiangular cases. We then focus on the particular case $p=2$. Section \ref{sec:Numerical solution} is devoted to the numerical techniques and presents the fully non-linear solutions of the effective equations, including the phase diagrams for highly non-uniform lattice black branes. We conclude in Section \ref{sec:discussion}.

\section{Setup}
\label{sec:Setup}
In the large D limit, the leading order metric solution of the dynamical black $p$-brane in $D=n+p+3$ is solved by~\cite{Emparan:2015gva,Emparan:2016sjk}
\begin{align}
ds^2 = - \left(1-\frac{m(t,x)}{r^{2n}}\right)dt^2+2 dt dr-\frac{2p_i(t,x)dt dx^i}{r^{2n}}
+ \fr{n} dx^i dx_i + r^2 d\Omega_{n+1}^2,\label{eq:mericsol}
\end{align}
where $i=1,\dots,p$. The brane dimension $p$ is assumed to be finite at large $D$, and then the metric is expanded in $1/n$ rather than $1/D$. The factor $1/n$ in front of $dx^i dx_i$ is crucial to capture the GL instability and its related dynamics on the black brane, since the typical wavelength of the GL mode is given by $1/\sqrt{D} \simeq 1/\sqrt{n}$ at large $D$~\cite{Sorkin:2004qq,Asnin:2007rw}.
The dynamical degrees of freedom on the horizon $m(t,x)$ and $p_i(t,x)$ follow the effective equations
\begin{eqnarray}
 &&\partial_t m - \partial^2 m = - \partial_i p^i\\
 &&\partial_t p^i - \partial^2 p^i = \partial_i m - \partial_j \left(\frac{p^i p^j}{m}\right),\label{eq:largedeftfull}
\end{eqnarray}
where $p_i$ behaves as a vector with respect to $\delta_{ij}$. By perturbing the uniform solution
\begin{align}
 m(t,x) = m_0 + \veps\, \hat{m}\, e^{\Omega t} \cos({\bfk}\cdot {\bf x}),\quad  p^i(t,x) =  \veps\, \hat{p}^i \,e^{\Omega t} \sin({\bfk}\cdot {\bf x}),
\end{align}
the dispersion relation $\Omega(\bfk) = |\bfk|-|\bfk|^2$ is easily obtained, which shows the uniform solution is dynamically stable for any $\bfk$ with $|\bfk|>k_{\rm GL}=1$.
For static solutions, the equations reduce to the soap bubble equation,
\begin{eqnarray}
 \partial^2 \ln m + \fr{2} ( \partial \ln m)^2+ \ln m = C,
\end{eqnarray}
where $C$ is an integration constant.
Introducing $\cR = \ln m$, we have
\begin{eqnarray}
 \partial^2 \cR + \fr{2} (\partial \cR)^2 + \cR = 0
 \label{eqn:soap_bubble}
\end{eqnarray}
where we set $C=0$ by the scaling degree of freedom $\cR \to \cR +C$.

\begin{figure}[t]
	\begin{center}
		\includegraphics[width=0.5\textwidth]{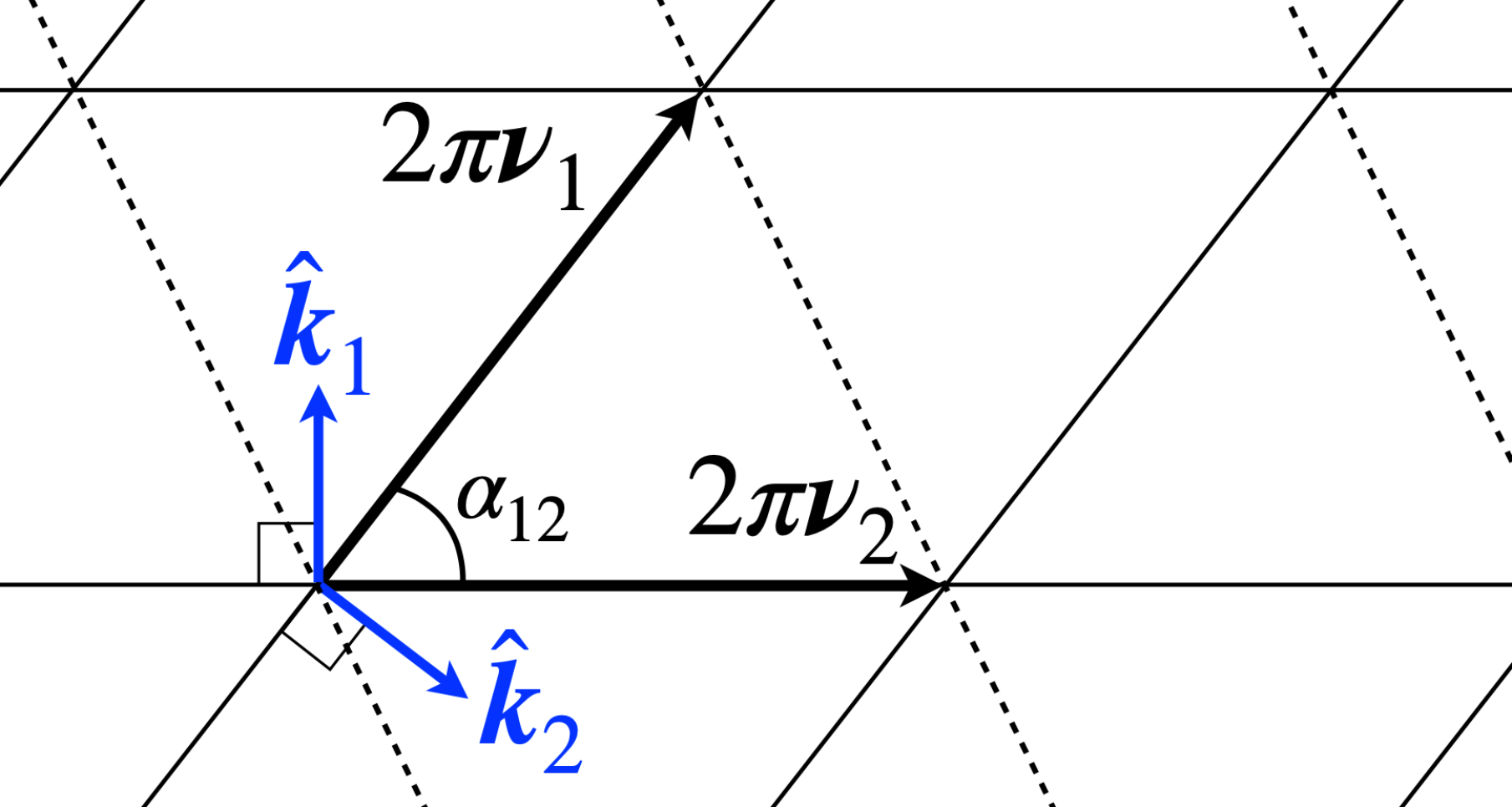}
		\caption{Unit wavenumber vectors $\{\hat{\bfk}_i\}$ and actual lattice period in the two-dimensional lattice. The dashed line represents the substructure formed by $2\pi (\bfnu_1-\bfnu_2)$.
		\label{fig:2dlattice-scheme} }
	\end{center}
\end{figure}
Now, we focus on the {\it black lattices}, \ie the static solutions with lattice structure. 
The lattice structure in ${\mathbb R}^p$ is characterized by $p$ independent vectors $\{ \bfk_i \}_{i=1,\dots,p}$,
with which the solution is written as
\begin{eqnarray}
\cR = \cR(\theta_1,\dots,\theta_p),\quad \theta_i = \bfk_i \cdot {\bf x}, 
\end{eqnarray}
with the identification $\theta_i \sim \theta_i + 2\pi$. 
If we choose another set of $p$ vectors $\{ \bfnu_i\}_{i=1,\dots,p}$ which satisfy $\bfk_i \cdot \bfnu_j = \delta_{ij}$,
the periodicity of the solution is expressed by
\begin{eqnarray}
 {\bf x} \rightarrow {\bf x} + 2\pi \sum_{i=1}^p n_i  \bfnu_i,\quad n_i\in {\mathbb Z}.
\end{eqnarray}
as illustrated in figure \ref{fig:2dlattice-scheme}.

\paragraph{Phase diagram}
Up to the leading order in $1/n$, the normalized total mass in a cell is given by
\begin{align}
 {\cal M} = \int_{\rm cell} e^{\cR(x)} d^px.\label{eq:def-normalizedmass}
\end{align}
However, at the leading order, the effective equation is scale invariant, and the solutions can scale to have arbitrary mass by $\cR(x) \to \cR(x)+C $. Thus, the solutions must be compared in scale invariant quantities.
One of such quantities is the mass normalized entropy~\cite{Andrade:2020ilm}
\begin{align}
 {\cal S}_1 = n\left(\frac{{\cal S}}{{\cal M}^{\frac{n+p+1}{n+p}}}-1\right) \simeq \frac{\delta {\cal S}}{{\cal M}}-\log {\cal M},
\end{align}
where the entropy difference from the total mass is given by the leading order solution
\begin{align}
 \delta {\cal S} = \int_{\rm cell} \left(-\fr{2} (\partial \cR)^2 + \cR\right) e^{\cR} dx^p.
\end{align}
However, by using the leading order equation~(\ref{eqn:soap_bubble}), it turns out this quantity vanishes when integrated over a period\footnote{A non-zero scaling parameter $C$ in eq.~(\ref{eqn:soap_bubble}) gives $\delta {\cal S} = C {\cal M}$, which simply adds a constant to ${\cal S}_1$.},
\begin{align}
 \delta {\cal S} = \int_{\rm cell} \left(-\fr{2} (\partial \cR)^2 + \cR\right) e^{\cR} dx^p = -\int_{\rm cell} \partial \cdot \left(e^\cR \partial \cR\right) dx^p=0.
\end{align}
And therefore, the mass normalized entropy is given by the total mass
\begin{align}
 {\cal S}_1 = - \log {\cal M}.\label{entropy-formula}
\end{align}
Note that this total mass cannot be scaled arbitrarily since the scaling is fixed in eq.~(\ref{eqn:soap_bubble}).
Particularly, for the uniform black brane $\cR(x)=0$, we have
\begin{align}
 {\cal M}_{\rm UBB} = V_{\rm cell},\quad  {\cal S}_{1,{\rm UBB}} = -\log V_{\rm cell},
\end{align}
where $V_{\rm cell}$ is the $p$-dimensional volume of a cell.

The Kaluza-Klein background also allows to define the tension related to the variation of the total mass with respect to the spatial boundary metric (see Appendix.~\ref{sec:brane-tension}),
\footnote{Ref.~\cite{Rozali:2016yhw} related the tension and quasi-local stress tensor with the wrong signature, and hence the conclusion is different.
Particularly, the minimum of the enthalpy is now given by the maximum of the bulk tension, which is dominated by the uniform solution. Therefore, we do not agree on the enthalpy.}
\begin{align}
{\cal T}^{ij} : =  - \fr{2} \int_{\rm cell} T^{ij} d^p x,\label{eq:def-tensionij}
\end{align}
where $T_{ij}$ is the quasi-local stress tensor of the effective theory. In the static configuration, $T_{ij}$ is given by
\begin{align}
 T^{ij} = e^{\cal R} (-\delta^{ij}+\partial^i \cR \partial^j {\cal R})+\partial^2 (e^\cR \delta^{ij}) - 2 \partial^i \partial^j e^\cR,
\end{align}
where $i,j$ are raised by $\delta_{ij}$. The last two terms vanish when integrated over a cell.
As opposed to the one dimensional case, the tension consists of multiple components corresponding to various changes in the lattice configuration.
To characterize the solution, we particularly focus on the {\it bulk tension}
\begin{align}
 {\cal T} := \fr{n} \delta_{ij} {\cal T}^{ij}= \fr{2n} \int_{\rm cell} e^\cR(p-(\partial \cR)^2) d^px = \fr{n}\int_{\rm cell} e^\cR(p/2-\cR) d^px,\label{eq:def-bulktension}
\end{align}
where in the second equality the effective equation~(\ref{eqn:soap_bubble}) is used with partial integration.
To eliminate the scale dependence, we rather use the mass-normalized tension
\begin{align}
 \tau := \frac{n{\cal T}}{\cal M} = \frac{p}{2} - \frac{\int_{\rm cell} \cR e^\cR d^px}{\int_{\rm cell} e^\cR d^px}.
 \label{eq:def-normalizedtension}
\end{align}
For the uniform black brane, we have
\begin{align}
 \tau_{\rm UBB} = \frac{p}{2}.
\end{align}
From the middle form in eq~(\ref{eq:def-bulktension}), it is obvious that the normalized tension reaches the maximum $\tau=p/2$ if and only if the solution is uniform.

\section{Perturbative analysis}
\label{sec:Perturbative analysis}

We start by studying the static perturbation around the uniform black brane $\cR(x)=0$,
\begin{eqnarray}
 \cR = \sum_{i=1}^\infty  \delta\cR_{i}.
\end{eqnarray}
At each order, the equation takes the form of
\begin{eqnarray}
( \partial^2 + 1 ) \delta\cR_i = S_i,
\end{eqnarray}
where $S_i$ is $i$-th order source term.
With the periodicity vectors $\{ \bfk_i \}_{i=1,\dots,p}$, the linear order solution is given by
\begin{eqnarray}
  \delta\cR_{1} = \veps \sum_i \lambda_i \cos \theta_i ,\quad \theta_i = \bfk_i \cdot {\bf x}. \label{eq:pertsol-1st-neq}
\end{eqnarray}
Here we expand for small $\veps$, and $\lambda_i$ determines the relative amplitudes between modes.
 The linear order equation requires each wavenumber to be on the threshold of the instability,
 \begin{align}
  k_i := |\bfk_i| = 1 + \ord{\veps}.
 \end{align}
As in the non-uniform black string analysis, we expect the wavenumber $k_i$ to be corrected by nonlinearities.

As observed in ref.~\cite{Dias:2017coo}, if a pair $(\bfk_1,\ \bfk_2)$ forms an equiangular lattice, \ie $\bfk_1\cdot \bfk_2 = -1/2+\ord{\veps}$, a special treatment is required. In this case, the triplet $(\bfk_1,\bfk_2, -\bfk_1-\bfk_2)$ forms an equilateral triangle, and hence one should respect the symmetry between the three, that is, $\cos(\theta_1+\theta_2)$ should be added to the linear solution. In higher dimensions, we will have more equiangular lattices. For example, a triplet $(\bfk_1,  \bfk_2, \bfk_3)$ will make a three dimensional equiangular lattice if they satisfy $\bfk_i\cdot \bfk_j = -1/3+\ord{\veps}$ for any pair, in which case the quartet $(\bfk_1,\bfk_2,\bfk_3,-\bfk_1-\bfk_2-\bfk_3)$ forms a regular tetrahedron and then, $\cos(\theta_1+\theta_2+\theta_3)$ joins the linear solution. These properties can be seen more clearly by examining higher order perturbations in the later section. 

Next, we study the lattices on general $p$-branes in the non-equiangular case by perturbative expansion. Then, we present more detailed results for the two dimensional lattice in both non-equiangular and equiangular cases.

\subsection{General analysis on non-equiangular lattice}
First, we consider general cases with non-equiangular configuration. We assume that for $q=2,3,\dots,p$ any set of $q$
different wavenumber vectors $(\bfk_{i_1},\dots,\bfk_{i_q})$
do not satisfy
\begin{align}
 \gamma_{i_{n}i_{m}} := -\hat{\bfk}_{i_n} \cdot \hat{\bfk}_{i_m} = 1/q \quad {\rm for} \ {\rm all} \  n\neq  m,
\end{align}
where $\gamma_{ij}$ denotes the minus cosine between two vectors\footnote{We added the minus sign because the angle in the momentum space $\bar{\alpha}_{ij}$ and the actual lattice angle $\alpha_{ij}$ are related by $\bar{\alpha}_{ij}=\pi-\alpha_{ij}$.}.
The second order source from eq.~(\ref{eq:pertsol-1st-neq}) is given by
\begin{eqnarray}
 S_2 = - \fr{2}(\partial  \delta\cR_1)^2 = -\frac{\veps^2}{4} \sum_{i,j} \lambda_i \lambda_j \gamma_{ij} \left[\cos(\theta_i+\theta_j)-\cos(\theta_i-\theta_j)\right].
\end{eqnarray}
This is easily integrated to give the second order solution
\begin{eqnarray}
 \delta \cR_2 = - \frac{\veps^2}{4} \sum_{i,j}\lambda_i \lambda_j \left( a^{(+)}_{ij}\cos(\theta_i+\theta_j)+a^{(-)}_{ij} \cos(\theta_i-\theta_j)\right)
\end{eqnarray}
where
\begin{eqnarray}
 a^{(\pm)}_{ij} = \frac{\gamma_{ij} }{2\gamma_{ij}\mp 1}.
\end{eqnarray}
As we mentioned, if any pair of periodicity vectors has the equiangular configuration, the perturbative expansion breaks down as $a^{(+)}_{ij}$ diverges.
This simply indicates that $\cos(\theta_i+\theta_j)$ should be promoted to the linear order.

Using the linear and second order solutions, the third order source is obtained as
\begin{align}
& S_3 =   -\partial  \delta\cR_2 \cdot \partial  \delta\cR_1\nonum
 &=\veps^3 \left[
 \fr{4}\sum_{i} \chi^{(2)}_{i}\lambda_i \cos\theta_i+\fr{8} \sum_{i,j,k} \lambda_i \lambda_j \lambda_k a^{(+)}_{ij}(\gamma_{ik}+\gamma_{jk}) \cos(\theta_i+\theta_j+\theta_k)\right.\nonum
&\quad \left.- \fr{8}\sum_{i,j} \sum_{k \neq i,j}\lambda_i \lambda_j \lambda_k \left(a_{ij}^{(+)} (\gamma_{ik}+\gamma_{jk})+2 a^{(-)}_{jk} (\gamma_{ik}-\gamma_{ij})\right)\cos(\theta_i+\theta_j-\theta_k)\right] \label{eq:pert-src-3rd-neq}
\end{align}
where
\begin{eqnarray}
 \chi^{(2)}_{i} &=&\sum_j \lambda_j^2\left[(1-\gamma_{ij})a^{(+)}_{ij} +  (1+\gamma_{ij})a^{(-)}_{ij}\right]-a^{(+)}_{ii}\lambda_i^2\nonum
 &=& \sum_{j\neq i} \frac{2\gamma_{ij}^2}{4\gamma_{ij}^2-1}\lambda_j^2+\fr{3}\lambda_i^2.
\end{eqnarray}
The first term in eq.~(\ref{eq:pert-src-3rd-neq}) is the source of the secular behavior, and hence should be absorbed into the parameter renormalization
\begin{eqnarray}
 k_i = 1-\frac{\veps^2}{8}\chi^{(2)}_i.
\end{eqnarray}
Thus, the third order solution becomes
\begin{eqnarray}
 \delta \cR_3 = \frac{\veps^3}{8}\sum_{i,j,k} a^{(++)}_{ijk} \lambda_i \lambda_j \lambda_k \cos(\theta_i+\theta_j+\theta_k)+ \frac{\veps^3}{8}\sum_{i,j} \sum_{k\neq i,j} a^{(+-)}_{ij,k} \lambda_i \lambda_j \lambda_k \cos(\theta_i+\theta_j-\theta_k)
 \nonum
\end{eqnarray}
where
\begin{eqnarray}
 a^{(++)}_{ijk} = \frac{a_{ij}^{(+)}(\gamma_{ki}+\gamma_{jk})+a_{ki}^{(+)}(\gamma_{jk}+\gamma_{ij})+a_{jk}^{(+)}(\gamma_{ki}+\gamma_{ij})}{6(\gamma_{ij}+\gamma_{jk}+\gamma_{ki}-1 )}
\end{eqnarray}
and
\begin{eqnarray}
 a^{(+-)}_{ij,k} = \frac{a^{(+)}_{ij}(\gamma_{ki}+\gamma_{jk})+a^{(-)}_{ki}\gamma_{kj}+a^{(-)}_{jk}\gamma_{ik}-\left(a^{(-)}_{jk}+a^{(-)}_{ki}\right)\gamma_{ij}}{2(\gamma_{ki}+ \gamma_{jk} -\gamma_{ij}+1 )}.
\end{eqnarray}
As in the second order, $a^{(++)}_{ijk}$ diverges if the triplet $(\bfk_i,\bfk_j,\bfk_k)$ is in the equiangular configuration, which indicates $\cos(\theta_i+\theta_j+\theta_k)$ should come to the linear order.
\cout{
If we define the symmetric parameters
\begin{eqnarray}
 {\sf s}_{ijk} = \gamma_{ij}+\gamma_{jk}+\gamma_{ki},\quad
 {\sf t}_{ijk} = \gamma_{ij}\gamma_{jk}+\gamma_{jk}\gamma_{ki}+\gamma_{ki}\gamma_{ij},\quad
{\sf u}_{ijk} = \gamma_{ij}\gamma_{jk}\gamma_{ki}
\end{eqnarray}
and partially symmetric parameters
\begin{eqnarray}
 {\sf s}_{ij,k} = \gamma_{ik}+\gamma_{jk},\quad  {\sf t}_{ij,k} = \gamma_{ik}\gamma_{jk},
\end{eqnarray}
the coefficients are written simpler
\begin{eqnarray}
 a^{(++)}_{ijk} = \frac{{\sf u}_{ijk}(4{\sf s}_{ijk}-3)+{\sf t}_{ijk}(1-{\sf s}_{ijk})}{2(1-2{\sf s}_{ijk}+4{\sf t}_{ijk}-8{\sf u}_{ijk})(1-{\sf s}_{ijk})},
\end{eqnarray}
\begin{eqnarray}
 a^{(+-)}_{ij,k} =\frac{ (4\gamma_{ij}-1){\sf t}_{ij,k} {\sf s}_{ij,k}-{\sf t}_{ij,k} (2\gamma_{ij}-1)^2-{\sf s}_{ij,k} \gamma_{ij} (2 \gamma_{ij}-{\sf s}_{ij,k}-1)}{(2\gamma_{ij}-1)(4{\sf t}_{ij,k}+2{\sf s}_{ij,k}+1)({\sf s}_{ij,k}-\gamma_{ij}+1)}.
\end{eqnarray}
}

\subsection{Two-dimensional lattice}
Now, we perform a more detailed analysis on the two-dimensional lattice, both in the non-equiangular and equiangular cases.
For simplicity, we only consider the equilateral configuration where two directions are symmetric
\begin{align}
 k := k_1=k_2,\quad \lambda_1=\lambda_2=1.
\end{align}
Then, the solution is characterized by the wavelength $k$ and the cosine of the lattice angle
\begin{align}
 \gamma :=  - \hat{\bfk}_1 \cdot \hat{\bfk}_2,\quad (0<\gamma<1).
\end{align}

\subsubsection{Non-equiangular lattice}
\label{sec:pert-2d-noneq}
First, we consider the non-equiangular case $\gamma\neq 1/2$.
Repeating the procedure in the previous section, we obtained the perturbative solution up to the fifth order in which the periodicity is determined up to $\ord{\veps^4}$ by
\begin{align}
& k = 1 - \frac{\veps^2}{24} \frac{10\gamma^2-1}{4\gamma^2-1}+\frac{\left(8800 \gamma ^8-16456 \gamma ^6+8232 \gamma ^4-1567 \gamma ^2+19\right) \veps^4}{6912 \left(\gamma ^2-1\right) \left(4 \gamma ^2-1\right)^3}.\label{eq:k-non-equiangular}
\end{align}
The mass and entropy are computed accordingly as
\begin{align}
 {\cal M}  = \left(1+\frac{(1-10\gamma^2)}{48(4\gamma^2-1)}\veps^4\right) \fr{k^2\sqrt{1-\gamma^2}}
\end{align}
and
\begin{align}
{\cal S}_1 = {\cal S}_{1,{\rm UBB}}(k(\veps),\gamma)+\frac{10\gamma^2-1}{48(4\gamma^2-1)}\veps^4,
\end{align}
where the entropy for the uniform black brane is given by
\begin{align}
 {\cal S}_{1,{\rm UBB}}(k,\gamma) =- \log \left(\frac{(2\pi)^2}{k^2\sqrt{1-\gamma^2}}\right).\label{eq:entropy-UBB}
\end{align}
The tension is also given by
\begin{align}
 \tau = 1-\frac{\veps^2}{2}+\frac{11-178\gamma^2+464\gamma^4}{72(1-4\gamma^2)^2}\veps^4.
\end{align}
Note that the physical quantities only have even powers of the amplitude $\veps$. This reflects the fact that the change of the signature $\veps \to -\veps$ gives the identical solution with the spatial translation $\theta_i \to \theta_i+\pi$. Thus, this branch is one-sided.

\subsubsection{Equiangular lattice}
Next, we consider the equiangular case. The periodicity vectors $\bfk_1,\bfk_2$ are such that $\hat{\bfk}_1\cdot \hat{\bfk}_2 = -1/2$ and $k_1=k_2=k$.
The linear order solution should have the symmetry with respect to the rotation by $60^\circ$, or symmetry between $\cos\theta_1,\cos\theta_2$ and $\cos(\theta_1+\theta_2)$,
\begin{equation}
 \delta \cR_1 = \veps ( \cos \theta_1+ \cos \theta_2+  \cos (\theta_1+\theta_2))\label{eq:pertsol-1st-eq}
\end{equation}
where
\begin{equation}
\theta_{i} := \bfk_i\cdot {\bf x},\quad k_1 = k_2 = k = 1+\ord{\veps}.
\end{equation}
The second order source is given by
\begin{align}
& S_2 =-\frac{\veps^2}{4}(\cos\theta_1+\cos\theta_2+\cos(\theta_1+\theta_2)) +\frac{\veps^2}{4} \left[-3+\cos(2\theta_1)+\cos(2\theta_2)+\cos(2\theta_1+2\theta_2)\right.\nonum
&\left.\hspace{5.5cm} +\cos(2\theta_1+\theta_2)+\cos(2\theta_2+\theta_1)+\cos(\theta_1-\theta_2) \right].
\end{align}
Here we see that the first three terms in the source are resonant and hence to be absorbed to the period
\begin{equation}
k = 1+\frac{\veps}{8}.
\end{equation}
Then, the second order solution becomes
\begin{align}
& \delta\cR_2 =  -\frac{\veps^2}{4} \left[3+\fr{3}\cos(2\theta_1)+\fr{3}\cos(2\theta_2)+\fr{3}\cos(2\theta_1+2\theta_2)\right.\nonum
&\left.\hspace{2cm} +\fr{2}\cos(2\theta_1+\theta_2)+\fr{2}\cos(2\theta_2+\theta_1)+\fr{2}\cos(\theta_1-\theta_2) \right].
\end{align}
Repeating the analysis up to $\ord{\veps^5}$, we obtain
\begin{align}
 k = 1 +\frac{1}{8}\veps-\frac{43}{384}\veps^2-\frac{427}{9216}\veps^3+\frac{24137}{884736}\veps^4
\end{align}
and
\begin{align}
{\cal M} = \left(1+\fr{16}\veps^3-\frac{7}{128}\veps^4-\frac{319}{3072}\veps^5\right)\frac{2}{\sqrt{3} k^2}.
\end{align}
Thus, the entropy and tension are given by
\begin{align}
 {\cal S}_1={\cal S}_{1,{\rm UBB}}(k(\veps),1/2) -\fr{16}\veps^3+\frac{7}{128}\veps^4+\frac{319}{3072}\veps^5
\end{align}
and
\begin{align}
 \tau = 1-\frac{3}{4}\veps^2-\frac{9}{16}\veps^3+\frac{329}{768}\veps^4+\frac{205}{288}\veps^5.
\end{align}
${\cal S}_{1,{\rm UBB}}$ is the entropy of the uniform black brane~(\ref{eq:entropy-UBB}).
As opposed to the non-equiangular case, the different signs of $\veps$ lead to distinct branches. This corresponds to the fact that the linear solution~(\ref{eq:pertsol-1st-eq}) cannot flip the entire sign only by the translation $\theta_i \to \theta_i + \pi$.

\section{Numerical solution}
\label{sec:Numerical solution}

In order to properly solve the soap bubble equation (\ref{eqn:soap_bubble}) with the suitable periodicity,
it is convenient to introduce oblique coordinates $(u, v)$, adapted to the lattice, defined as
\begin{align}
 u := \theta_1 = k(\sqrt{1-\gamma^2} x - \gamma y),\quad v:=\theta_2 = ky\, ,
\end{align}
where ${\bf x}=(x,y)$ are the Cartesian coordinates. The wavenumber vectors are given by
 \begin{align}
\bfk_1 = (k\sqrt{1-\gamma^2},-k\gamma),\quad \bfk_2 = (0,k).
\end{align}
Inversely, we have
\begin{equation}
x = \frac{u + \gamma v}{k \sqrt{1-\gamma^2}}, \quad y = \frac{v}{k}\, .
\label{eqn:uv_coords}
\end{equation}
\begin{figure}[h]
	\begin{center}
		\includegraphics[width=0.5\textwidth]{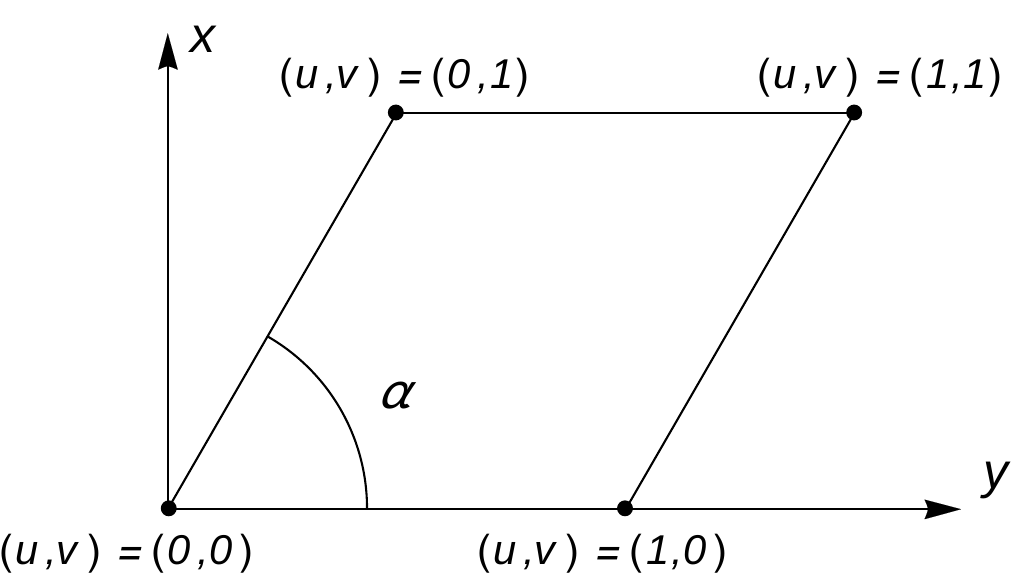}
		\caption{Unit cell of the lattice, with $\gamma = \cos \alpha$, as defined by (\ref{eqn:uv_coords}).\label{fig:uv_coords}}
	\end{center}
\end{figure}
In these coordinates, the periodicity of the solution becomes as simple as 
\begin{equation}
(u, v) \to (u, v) + 2\pi (n_u, n_v), \quad (n_u, n_v) \in {\mathbb Z}^2,
\end{equation}
as visualized in Figure \ref{fig:uv_coords}.
The soap bubble equation is then solved by the standard Newton-Raphson method on a square two-dimensional $N \times N$ Fourier grid with spectral differentiation matrices following the procedure in \cite{Dias:2015nua}. Particularly, we represent $\RR$ as a column vector of length $N^2$ by co-lexicographic ordering of its values at the collocation points and iterate 
\begin{equation}
\RR_{n+1} = \RR_n - \Delta_n
\end{equation}
until the desired precision. Here $\Delta_n$ is the solution (at the $n$-th iteration) of the linear system 
\begin{equation}
J \cdot \Delta = F,
\end{equation}
where $(\cdot)$ denotes matrix multiplication and 
\begin{equation}
F = \RR + k^2 \left[\partial^2 \RR + \frac12(\partial \RR)^2 - \gamma ( \partial_u \RR \partial_v \RR + 2\partial_v \partial_u \RR )\right],
\end{equation}
\begin{equation}
\begin{split}
J = I &+ k^2 \left[ D_u^2 + D_v^2 + \text{diag}(\partial_u \RR) \cdot D_u + \text{diag}(\partial_v \RR) \cdot D_v \right] \\
 & - k^2 \gamma \left[   \text{diag}(\partial_u \RR) \cdot D_v + \text{diag}(\partial_v \RR) \cdot D_u + D_v \cdot D_u  \right],
\end{split}
\end{equation}
with $D_u$ and $D_v$ the pseudospectral differentiation matrices with respect to coordinates $u$ and $v$ respectively. The initial guess for the iterative method can be taken as 
\begin{equation}
\RR_0 = A \cos u + B \cos v + C \cos(u + v).
\end{equation}
The value of the constants $A, B, C$ can be varied to select the branch of solutions the where we want the method to converge. A resolution of $N=20$ is often sufficient, although solutions at small values of $k$ or large values of $\gamma$ require more resolutions.

\subsection{Equiangular lattice}

\begin{figure}[H]
	\begin{center}
		\includegraphics[width=\textwidth]{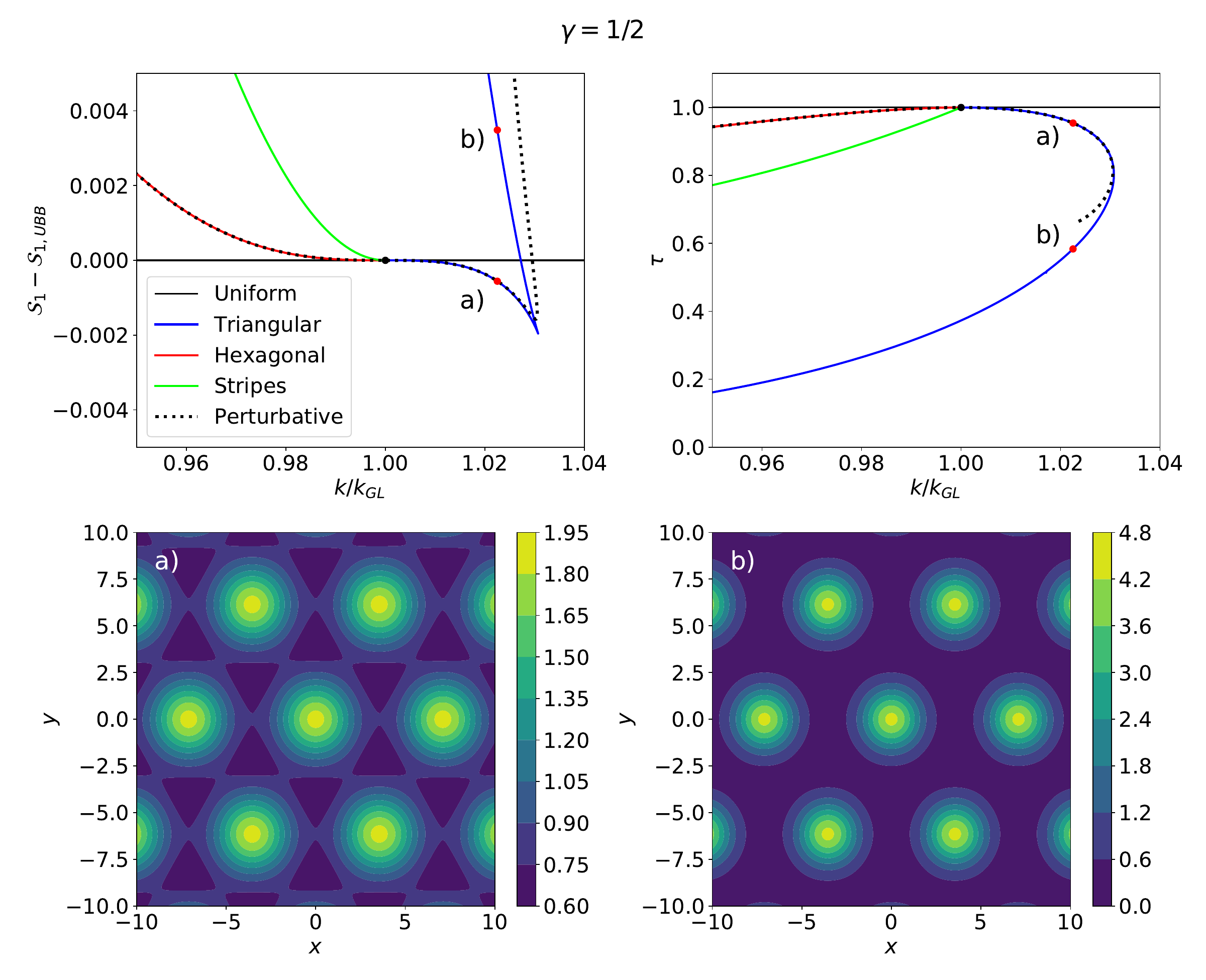}
		\caption{Entropy $\cS_1$ (top left) and tension $\tau$ (top right) for the equiangular lattice. The triangular branch presents a cusp at $k \approx 1.03\, k_{GL}$, and becomes entropically favored over the uniform brane at $k \approx 1.027\, k_{GL}$. The bottoms are the mass density plots showing the change between the triangular equiangular lattices before (a) and after (b) the cusp.
		 \label{fig:Equiangular_S1} }
	\end{center}
\end{figure}

In the equiangular case ($\gamma = 1/2$) we have a two-sided branch originating at the Gregory-Laflamme point $k_{GL}$ from the mode (\ref{eq:pertsol-1st-eq}). We refer to these two sides as {\it hexagonal} and {\it triangular} branches for $\veps < 0$ and $\veps > 0$ respectively. The numerical solution for both branches at $k = 0.95\, k_{GL}$ are shown in Figure \ref{fig:Equiangular_heatmap} as heat maps. In the phase diagram of the entropy $\cS_1$, we observe the presence of a cusp in the triangular branch (Figure \ref{fig:Equiangular_S1}), where both $k$ and $\cS_1$ simultaneously reach a maximum. The triangular phase is entropically favored over the uniform brane for $k < 1.027\, k_{GL}$.

Additionally, we obtain three one-dimensional branches, which we call {\it black stripes}. They are branching from the modes $\RR \sim \cos u$, $\RR \sim \cos v$ and $\RR \sim \cos(u+v)$ separately. These phases are equivalent under rotations of 60$^\circ$, and hence have an identical phase diagram bifurcating from $k=k_{\rm GL}$.

\subsection{Non-equiangular lattice}

\begin{figure}[H]
	\begin{center}
		\includegraphics[width=\textwidth]{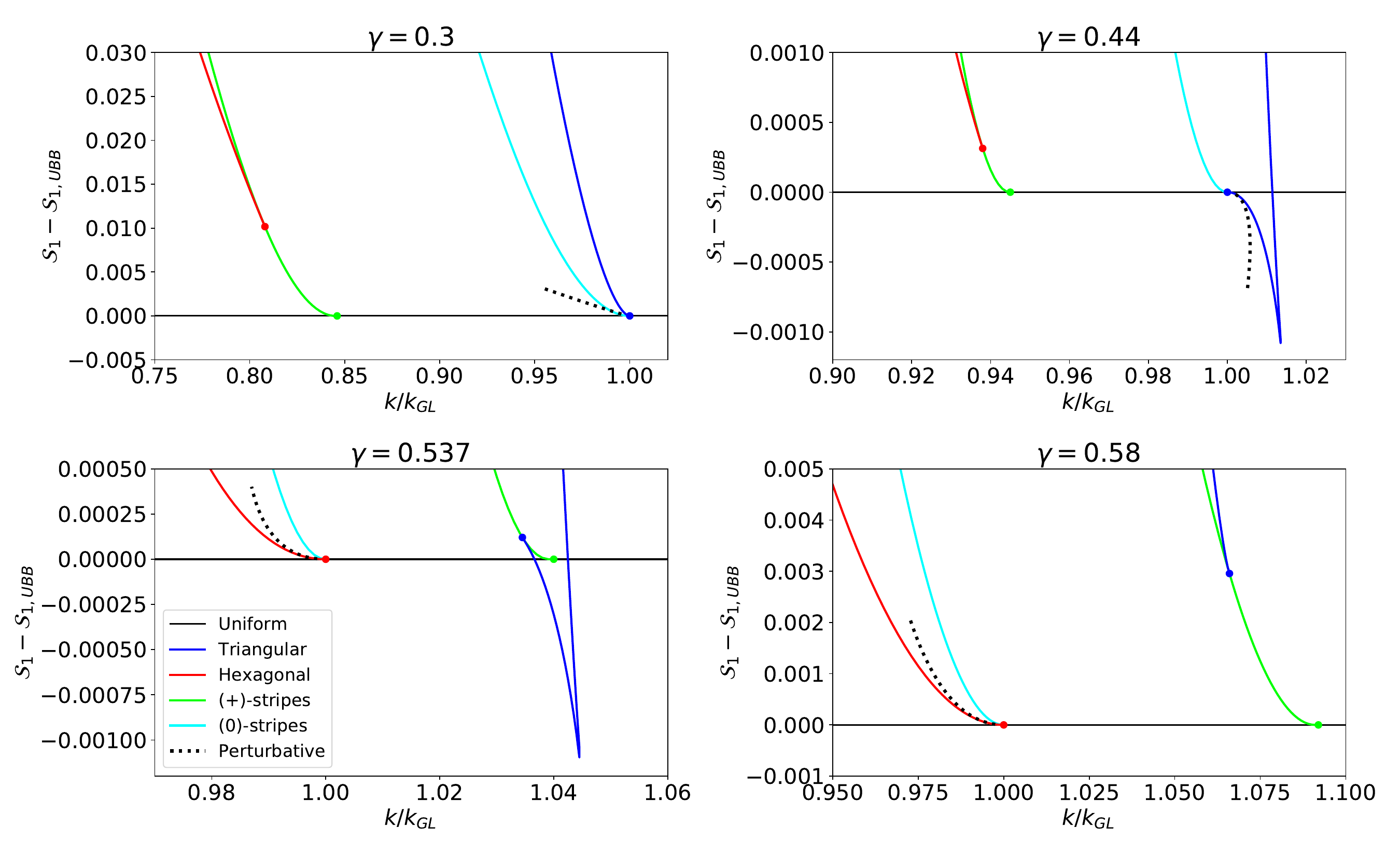}
		\caption{Entropy phase diagrams for non-equiangular lattice above and below $\gamma = 1/2$. \label{fig:NonEquiangularBranchingPoints_S1} }
	\end{center}
\end{figure}

\begin{figure}[H]
	\begin{center}
		\includegraphics[width=\textwidth]{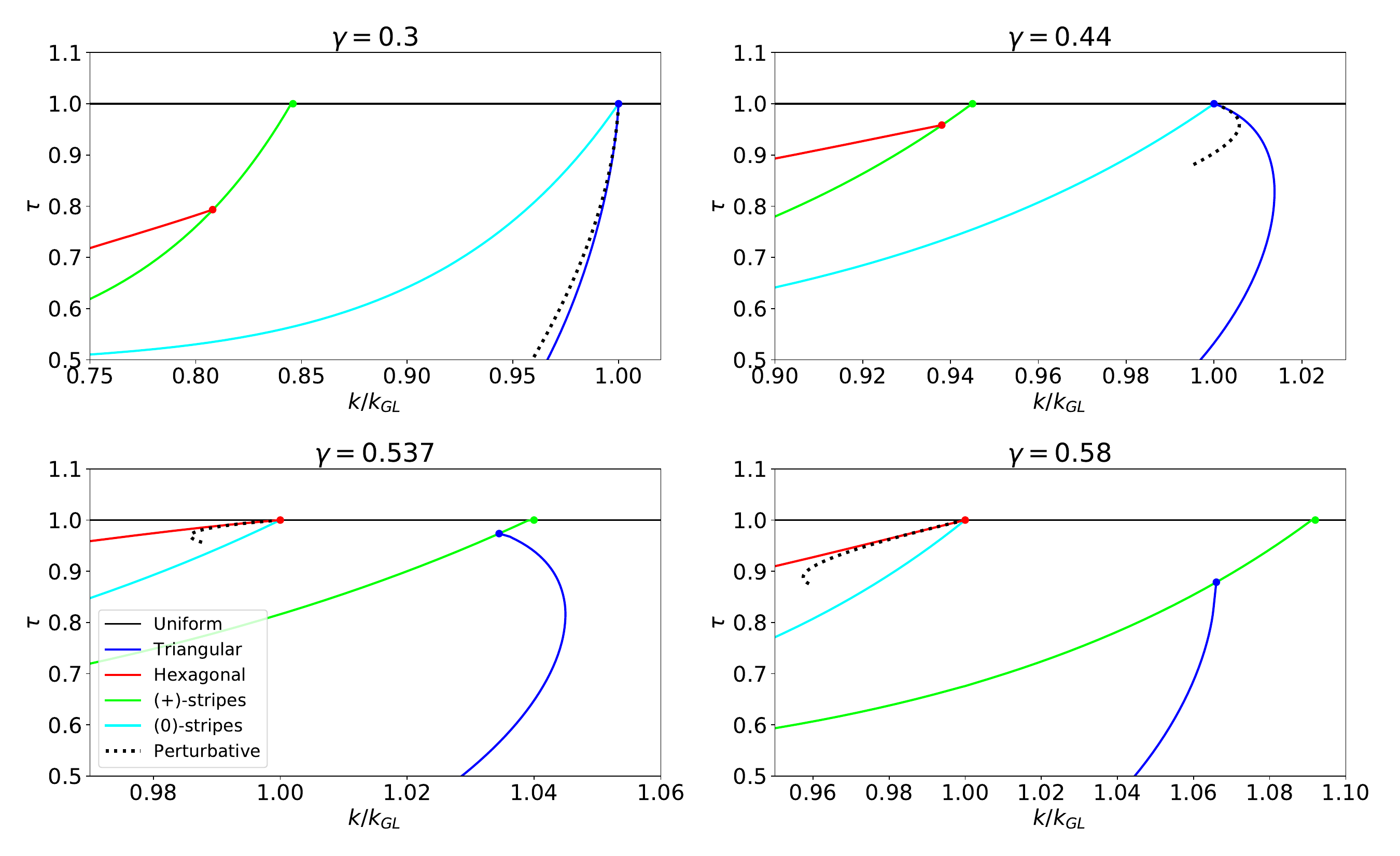}
		\caption{Tension phase diagrams for non-equiangular lattice above and below $\gamma = 1/2$. \label{fig:NonEquiangularBranchingPoints_tau} }
	\end{center}
\end{figure}

For $\gamma \neq 1/2$, as seen in the perturbative analysis in section~\ref{sec:pert-2d-noneq}, only one non-uniform branch with 2-dimensional dependence emerges from the Gregory-Laflamme point at $k = k_{GL}$. Namely, the triangular branch for $\gamma < 1/2$ and the hexagonal branch for $\gamma > 1/2$. The black stripes now split into two types. The modes $\RR \sim \cos u$ and $\RR \sim \cos v$ give rise separate phases, that are equivalent under the transformation $u \leftrightarrow v$, which together we will call (0)-stripes. The $\RR \sim \cos(u+v)$ zero mode now appears at $k = k_{GL}/\sqrt{2(1-\gamma)}$, as shown in Figures \ref{fig:NonEquiangularBranchingPoints_S1}, \ref{fig:NonEquiangularBranchingPoints_tau}, originating a third branch of black stripes, that we will call (+)-stripes. This branch contains a zero mode on its own, where the ``remaining'' 2-dimensional branch starts, \ie the hexagonal branch for $\gamma < 1/2$ and the triangular branch for $\gamma > 1/2$. In this case, the translational symmetry of the black brane is broken along two orthogonal directions in two steps: First in the $u+v$ direction, and then in $u-v$. Only at $\gamma = 1/2$, both triangular and hexagonal branches merge in a single, two-sided branch.

Interestingly, the cusp in the triangular branch observed for $\gamma=1/2$ (Figure \ref{fig:Equiangular_S1}) only exists for a finite range of the angle parameter $\gamma$, namely between two critical values that we will call $\gamma_1$ and $\gamma_3$.

\begin{figure}[H]
	\begin{center}
		\includegraphics[width=\textwidth]{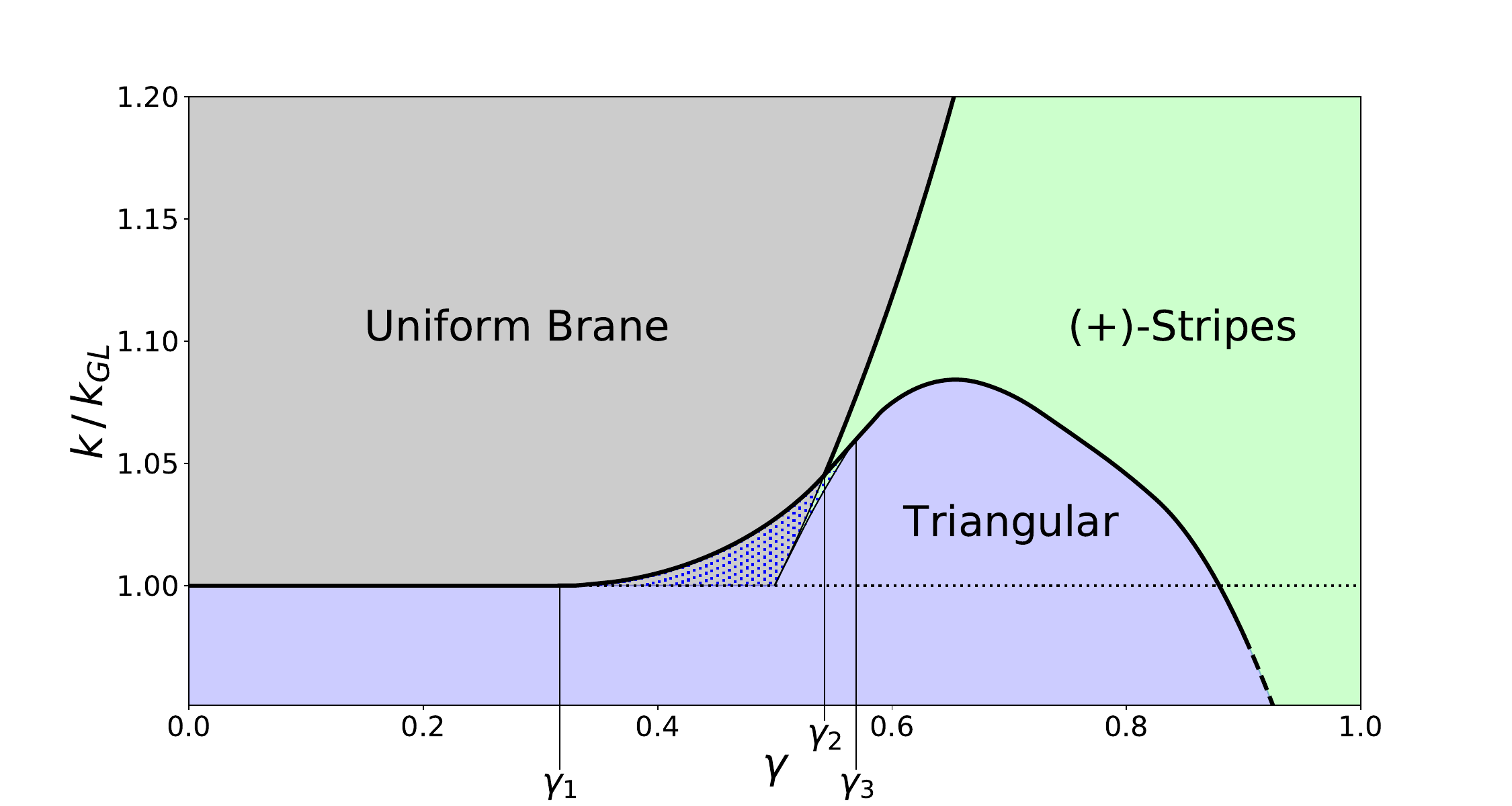}
		\caption{Thermodynamically preferred phases in the ($\gamma, k$) plane bordered with thick curves. The triangular phase reaches the maximum of $k$ at $(\gamma_{\rm max},k_{\rm max})\approx (0.65,1.084k_{\rm GL})$. At $\gamma \to 1$, the $(+)$-stripes become dominant. The dotted regions over gray and green backgrounds depict the presence of metastable phases of uniform branes and (+)-stripes, respectively. The curve beyond $\gamma = 0.9$ is an extrapolation.
		 \label{fig:stablephase_numerical} }
	\end{center}
\end{figure}

From eq.~(\ref{eq:k-non-equiangular}) one can derive that $\gamma_1 = 1/\sqrt{10} \approx 0.32$, where the coefficient of $\veps^2$ flips sign. The value of $\gamma_3$ has been numerically approximated as $\gamma_3 \approx 0.57$. Additionally, we define another critical angle, $\gamma_2 \approx 0.54$, where the triangular phase intersects the branching point of the (+)-stripes in ($k, \cS_1$)-plane (see the lower panels in Figure~\ref{fig:NonEquiangularBranchingPoints_S1}). In other words, for $\gamma > \gamma_2$ there exists a segment of stable black stripes. We summarize the thermodynamically preferred phases for given lattice parameters $(\gamma,k)$ in Figure~\ref{fig:stablephase_numerical}. 

It is important to mention that the cusp in the triangular branch causes the presence of metastable states in the phases of uniform branes and (+)-stripes it bifurcates from. This happens at values of $k$ where the parent branch is dynamically stable but the triangular branch has a larger entropy.

\begin{figure}[H]
	\begin{center}
		\includegraphics[width=\textwidth]{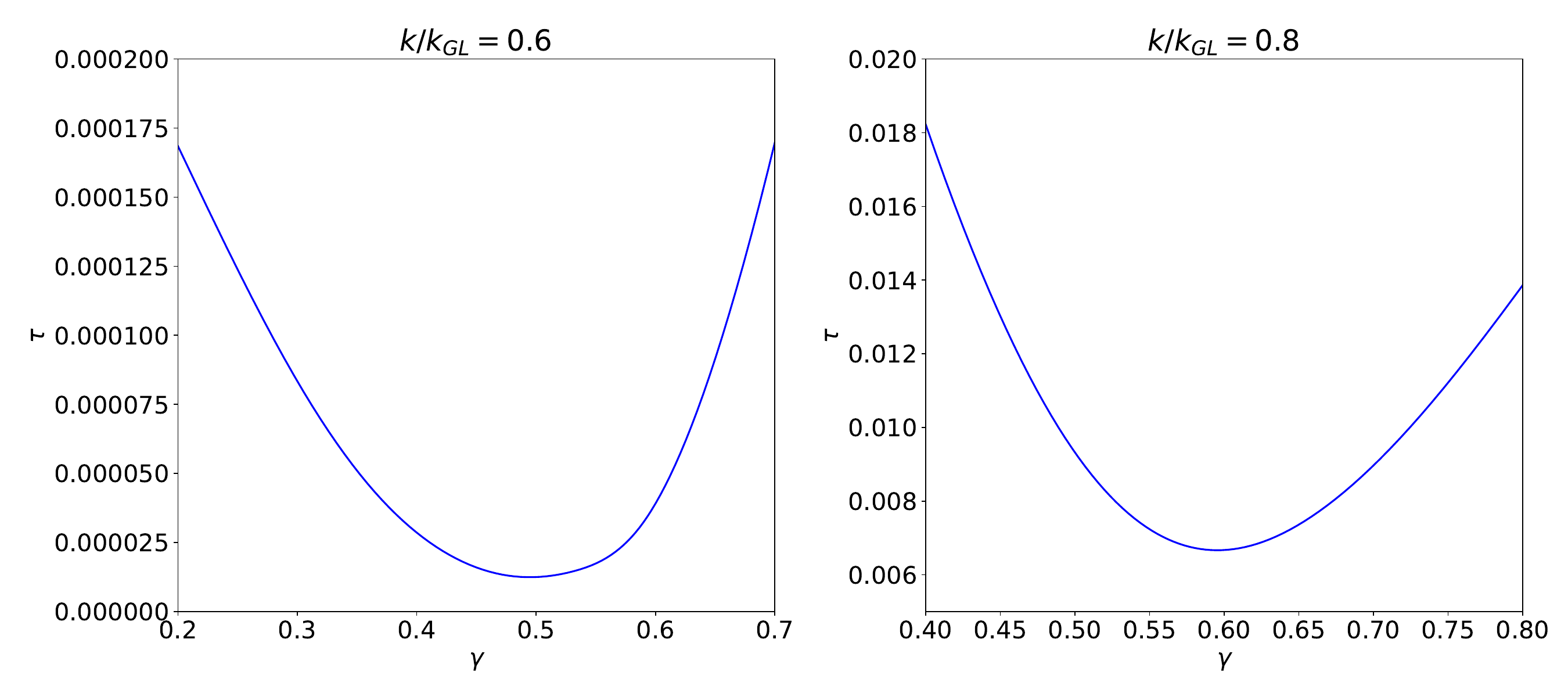}
		\caption{Tension as a phase of the angle parameter $\gamma$, for two fixed values of $k$. The value of $\gamma$ that minimizes $\tau$ is an increasing function of $k$. \label{fig:AngleDependence} }
	\end{center}
\end{figure}

We also plot the angular dependence of the tension in Figure~\ref{fig:AngleDependence}, which shows that it has a minimum for a certain value of $\gamma$. This is a similar behavior to the one observed in \cite{Rozali:2016yhw}.

\subsubsection{Branching from stripes}

\begin{figure}[H]
	\begin{center}
		\includegraphics[width=\textwidth]{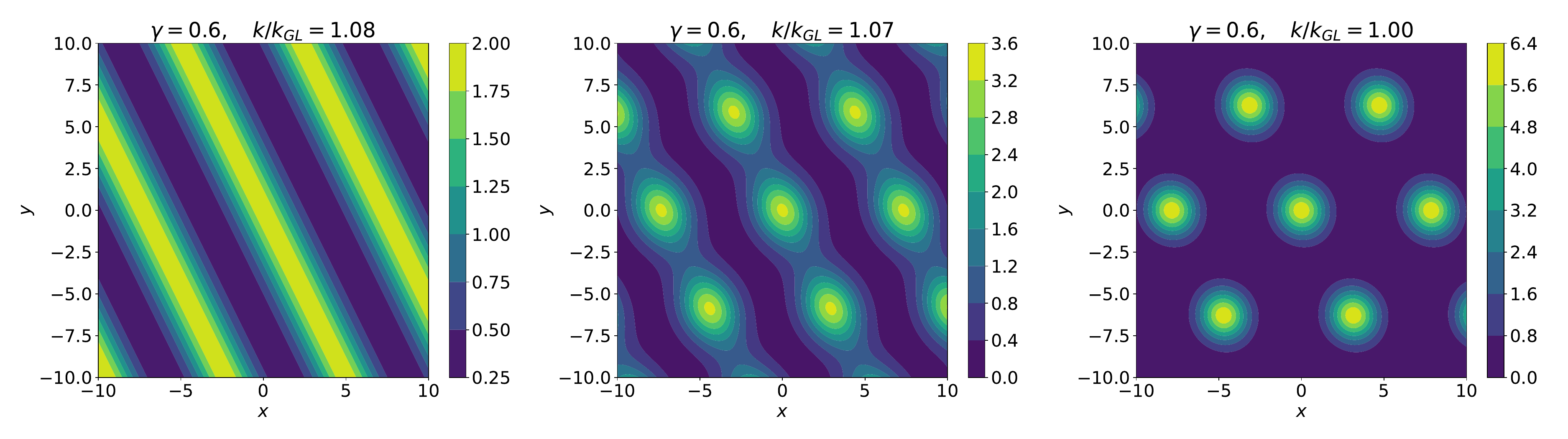}
		\caption{Transition between parallel (+)-stripes and the triangular lattice at $\gamma = 0.6$. The branching point happens at $k \approx 1.08 k_{GL}$.
\label{fig:Triangular_transition} }
	\end{center}
\end{figure}

\begin{figure}[H]
	\begin{center}
		\includegraphics[width=\textwidth]{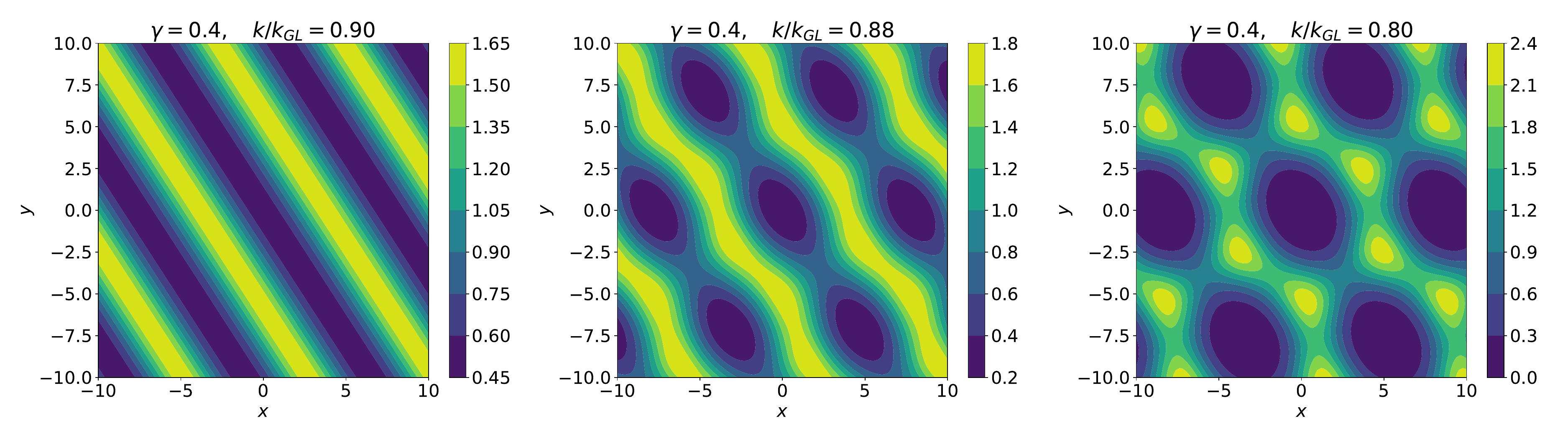}
		\caption{Transition between parallel (+)-stripes and the hexagonal lattice at $\gamma = 0.4$. The branching point happens at $k \approx 0.90 k_{GL}$.
\label{fig:Hexagonal_transition} }
	\end{center}
\end{figure}

Figures \ref{fig:Triangular_transition} and \ref{fig:Hexagonal_transition} show the transition from the (+)-stripes to the triangular and hexagonal lattices respectively. One can see that each transition takes place in a different way. For $\gamma>1/2$, the stripes start to fragment into isolated blobs. For $\gamma<1/2$, on the contrary, the stripes start to stick together forming bridges.

\subsection{Asymptotic phases at large deformation}

\begin{figure}[H]
	\begin{center}
		\includegraphics[width=\textwidth]{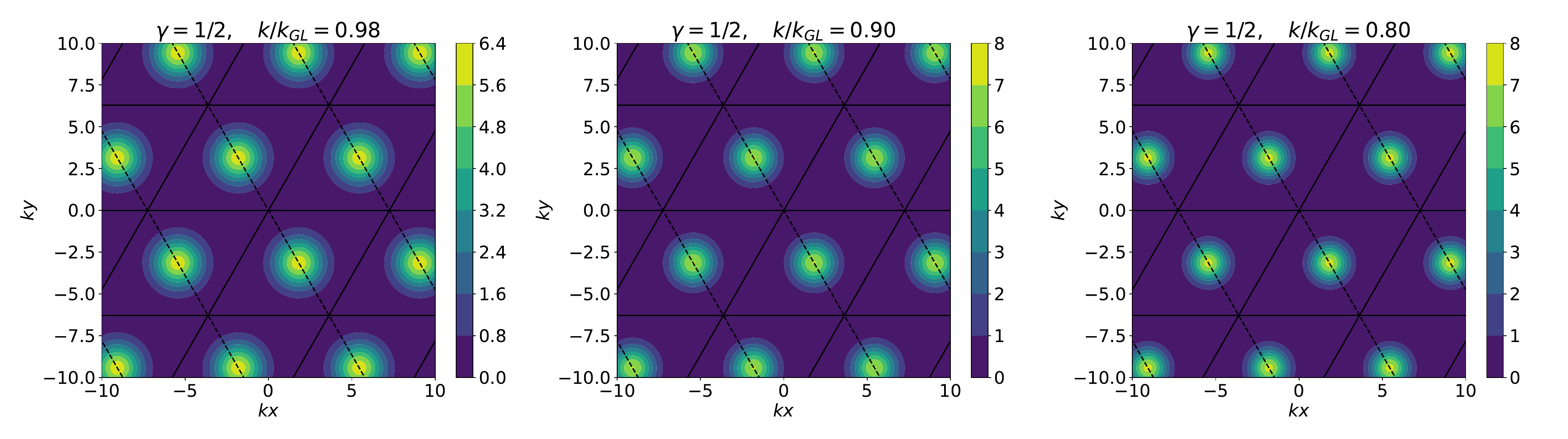}
		\caption{For small $k$, the triangular lattice approaches a lattice of isolated Gaussian blobs, with one blob per unit cell.
\label{fig:Triangular_asymptotics} }
	\end{center}
\end{figure}

\begin{figure}[H]
	\begin{center}
		\includegraphics[width=\textwidth]{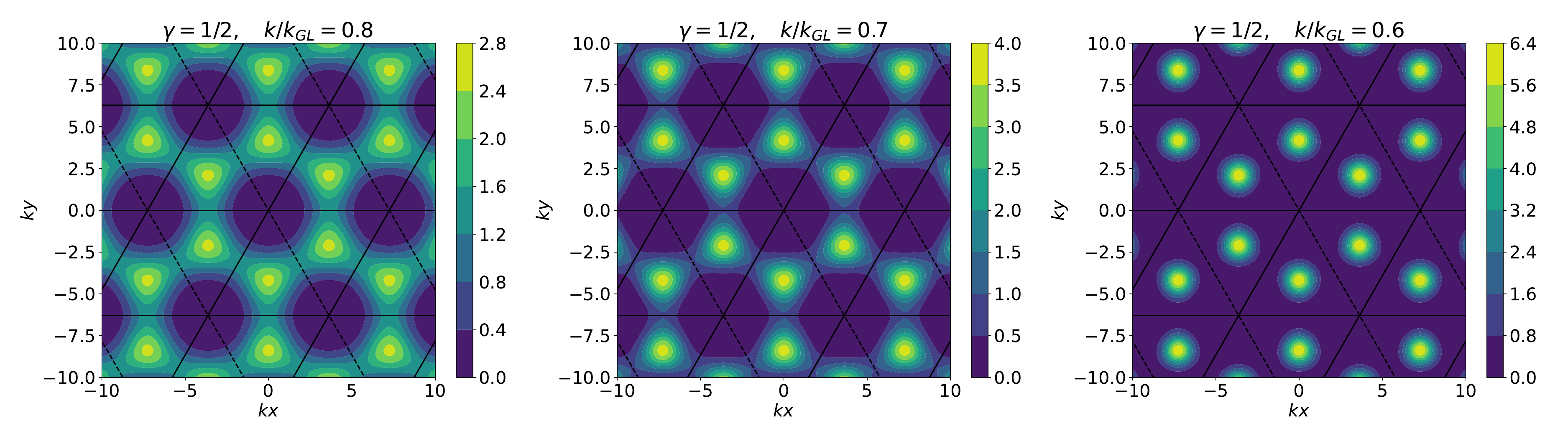}
		\caption{For small $k$, the hexagonal lattice approaches a lattice of isolated Gaussian blobs, with two blobs per unit cell.
\label{fig:Hexagonal_asymptotics} }
	\end{center}
\end{figure}

As $k \to 0$ both non-uniform phases with 2-dimensional dependence (triangular and hexagonal) asymptote to a lattice of Gaussian blobs of the form described in \cite{Andrade:2018nsz}. The evolution of such phases as we take smaller values of $k$ is depicted in Figures \ref{fig:Triangular_asymptotics}, \ref{fig:Hexagonal_asymptotics} as heat maps. 

\begin{figure}[H]
	\begin{center}
		\includegraphics[width=\textwidth]{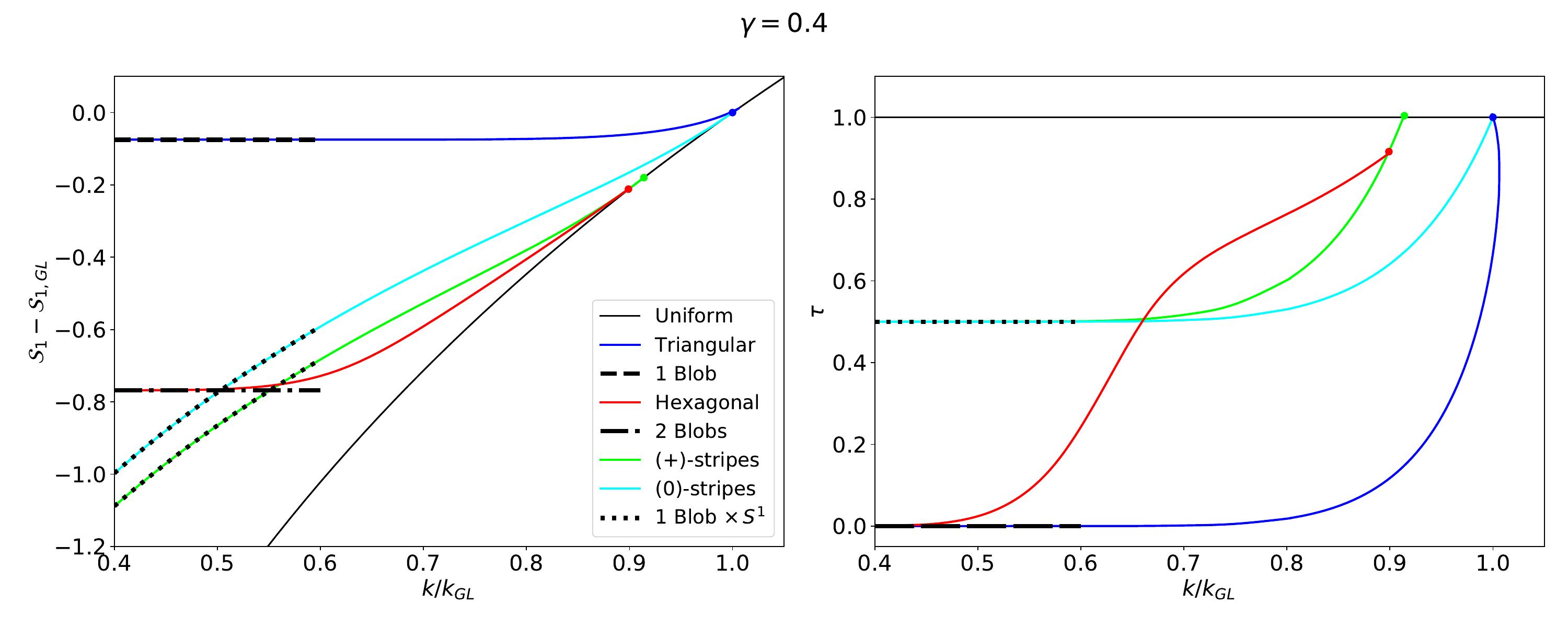}
		\caption{ At low values of $k/k_{GL}$ both triangular and hexagonal lattices approach a lattice of Gaussian blobs. One blob per unit cell in the triangular case, and two blobs per unit cell in the hexagonal case. The black stripes approach a Gaussian section as well.
\label{fig:Lowk_S1} }
	\end{center}
\end{figure}

The two phases, however, differ in the number of blobs per unit cell: The triangular phase has a single blob, while the hexagonal phase ends up having two blobs per unit cell. This results in $\mathcal{S}_1$ approaching two different values for small $k$ (see Figure \ref{fig:Lowk_S1}). The limiting values at $k\to0$ are easily estimated by approximating the solution as a group of isolated Gaussian blobs.
A blob centered at $(x_0,y_0)$ in Cartesian coordinates is written by
\begin{align}
 e^{\cR(x,y)} \simeq e^{2-\fr{2}((x-x_0)^2+(y-y_0)^2)}.
\end{align}
Then, $q$-blobs in a cell make the mass
\begin{align}
 {\cal M} \simeq q \times \int_{{\mathbb R}^2} e^{2-\fr{2}(x^2+y^2)} dxdy = 2\pi e^2 \, q,
\end{align}
and the entropy is given by
\begin{align}
 \cS_1 \simeq -\log (2\pi e^2 q).
\end{align}
The tension is also evaluated as
\begin{align}
 \tau \simeq 1-\frac{2\pi e^2 q}{2\pi e^2 q} =0.
\end{align}
In a similar way, the limiting phase of black stripes is approximated by the direct product of a Gaussian blob and $S^1$, by which one can expect the topology changing transition to parallel black strings.
For example, $(+)$-stripes is approximated by 
\begin{align}
 e^{\cR(u,v)} \simeq e^{1-\fr{4k^2(1-\gamma)}(u+v)^2}.
\end{align}
This leads to
\begin{align}
 {\cal M} \simeq \fr{k^2\sqrt{1-\gamma^2}}\int^{2\pi}_0 dv \int_{\mathbb R} e^{1-\fr{4k^2(1-\gamma)}(u')^2} du' = \frac{4\pi^{3/2} e}{k\,\sqrt{1+\gamma}},
\end{align}
and hence
\begin{align}
 {\cal S}_1 \simeq -\log\left( \frac{4\pi^{3/2} e}{k\,\sqrt{1+\gamma}}\right).
\end{align}
The entropy of $(0)$-stripes is also estimated by
\begin{align}
 {\cal S}_1 \simeq  -\log\left(\frac{2\sqrt{2}\pi^{3/2} e}{k\sqrt{1-\gamma^2}}\right).
\end{align}
The tension of both branches approaches the same value
\begin{align}
\tau \simeq \frac{1}{2}.
\end{align}

\section{Discussion}
\label{sec:discussion}

In this paper we have explored the phase space of static lattice deformations of black branes in the large $D$ limit. We focus on the cases where the primitive translation vectors have equal magnitude, \ie where the unit cell of the lattice is a rhombus. Among these cases we distinguish between two different scenarios, whether the unit cell has an angle of 60$^\circ$ (equiangular) or some other value (non-equiangular).

In the equiangular case ($\gamma = 1/2$) we obtain two distinct phases of lattices that branch off from the two signs of excitation of the same zero mode, $\RR \sim \cos u + \cos v + \cos(u+v)$, at $k = k_{GL}$. These two phases have triangular and hexagonal structures, respectively, as observed in the $D=6$ case~\cite{Dias:2017coo}. 

Interestingly, the triangular phase presents a cusp in its phase diagram, which makes the triangular branch thermodynamically preferred over the uniform branch slightly beyond the zero mode wavenumber ($k_{\rm GL} < k < k^*_{1/2} \approx 1.027k_{GL} $), while the uniform branch dominates for shorter periods. This is not the case in $D=6$, where no cusp has been observed and the triangular branch becomes stable simply for $k<k_{\rm GL}$, indicating the existence of a critical dimension between $6<D<\infty$. We have also obtained a set of one-dimensional phases which correspond to deformations in a striped pattern, {\it black stripes}. These three deformations are equivalent under 60$^\circ$ rotations: $\RR \sim \cos u$, $\RR \sim \cos v$ and $\RR \sim \cos(u+v)$.

In the non-equiangular cases, the zero mode $\RR \sim \cos u + \cos v + \cos(u+v)$ splits in two. The mode $\RR \sim \cos u + \cos v$ still starts at $k = k_{GL}$, but now it gives rise to only one of 2-dimensional phases depending on the value of $\gamma$, \ie the triangular phase (for $\gamma < 1/2$) or the hexagonal phase (for $\gamma > 1/2$). The other phase no longer branches off from the uniform solution directly, but via the black stripes from the $\cos(u + v)$-mode, $(+)$-stripes, which comes out from the uniform solution at $k=k_{\rm GL}/\sqrt{2(1-\gamma)}$. The modes $\RR \sim \cos u$ and $\RR \sim \cos v$, $(0)$-stripes, create other phases of black stripes equivalent under $u \leftrightarrow v$. We have found that the phase of $(+)$-stripes becomes thermodynamically favored in a certain range of the parameter for $\gamma>\gamma_2 \approx 0.54$.
The $(+)$-stripes are going to dominate the stable phase as the lattice becomes narrower ($\gamma\to 1$).

We have observed the appearance of the cusp in the triangular phase for $\gamma_1=1/\sqrt{10} < \gamma < \gamma_3\approx 0.57$. As in the equiangular case, the cusp makes the triangular phase thermodynamically preferred slightly beyond the zero mode of the uniform brane $(k_{\rm GL}<k<k^*(\gamma))$. The extended upper bound $k^*(\gamma)$ is determined by the intersection between the triangular and uniform branches $(\gamma<\gamma_2)$ or triangular and $(+)$-stripes branches $(\gamma>\gamma_2)$ in the entropy plot.
In the extended domain of the stable triangular phase, the uniform brane and $(+)$-stripes are still dynamically stable above the onset of their instability, which makes them metastable states. Here we note that similar metastable phases are observed in the large $D$ effective theory on AdS black strings~\cite{Emparan:2021ewh}.

At large deformations, we have found that both the triangular and the hexagonal lattices asymptote to a periodic distribution of the Gaussian blobs described in \cite{Andrade:2018nsz}. However, the triangular lattice leads to a single blob per unit cell, while the hexagonal lattice contains two blobs per unit cell, thus leading to a different value of the mass-normalized entropies. The black stripes end up as an array of one dimensional Gaussian blobs, extended in the perpendicular direction.
In ref.~\cite{Suzuki:2020kpx}, the asymptotic phases of black strings and other rigidly rotating solutions~\cite{Licht:2020odx} can be expressed as the expansion of the large distance between separate blobs, by assuming the solution as a linear combination of basic blobs and glueing thin necks. One can try to find the analytic expansions at the large deformation limit by using this blob and neck construction.

There are several possible extensions of this work. Having a more general setup with non-equal periods, or more brane dimensions, are straightforward extensions. The dynamical evolution of the brane in the same lattice setup would also be an interesting topic to explore. One could use the large $D$ effective equations to find the endpoint of the GL instability as in the black string case~\cite{Emparan:2015gva}. With more than two brane dimensions, the large $D$ effective theory admits turbulence~\cite{Rozali:2017bll}. Therefore, one can expect an interplay between the GL instability and turbulence.

Solving higher order corrections in the $1/D$-expansion will also be informative. Usually, largely deformed solutions at the leading order theory are expected to break down when the minimum of mass density reaches $\sim e^{-D}$, where the topology-changing transition would take place.
At large enough deformations, the triangular and hexagonal phases can be connected to a black hole lattice, while the black stripes lead to parallel black strings. Since the evolution of the hexagonal phase consists of two separate stages (Figure~\ref{fig:Hexagonal_asymptotics}), its topology-changing transition could be more complicated, that is, the black brane can first experience a transition to a lattice of holes punched in it as in bumpy Myers-Perry black holes to black rings~\cite{Emparan:2014pra}, and then fragment into a lattice of black holes.
The branching from the black stripes to the triangular or hexagonal branches can also lead to multiple topology-changing scenarios, depending on where the pinch-off occurs, before or after the branching.
To unravel the details of these topology changes, a fully-numerical analysis at finite dimensions would also be a viable option.

\section*{Acknowledgments}

This work is supported by ERC Advanced Grant GravBHs-692951, MICINN grant PID2019-105614GB-C22, AGAUR grant 2017-SGR 754, and State Research Agency of MICINN through the ``Unit of Excellence Mar\'ia de Maeztu 2020-2023'' award to the Institute of Cosmos Sciences (CEX2019-000918-M).
DL is supported by a Minerva Fellowship of the Minerva Stiftung Gesellschaft fuer die Forschung mbH.
RL has been supported by FCT/Portugal through the project IF/00729/2015/CP1272/CT0006, and by Next Generation EU though a Margarita Salas grant from the Spanish Ministry of Universities under the {\it Plan de Recuperaci\'on, Transformaci\'on y Resiliencia}.
RS is supported by JSPS KAKENHI Grant Number JP18K13541 and partly by Osaka City University Advanced Mathematical Institute (MEXT Joint Usage/Research Center on Mathematics and Theoretical Physics). 

\appendix
\section{Brane tension}\label{sec:brane-tension} 
Here we make manifest the relation between the mass and brane tension.
The quasi-local stress tensor is defined for the metric solution~(\ref{eq:mericsol}) by
\begin{align}
{\mathbf{T}}_{\mu\nu} = \lim_{r\to\infty} \frac{\Omega_{n+1} r^{n+1}}{8\pi G} (K g_{\mu\nu} - K_{\mu\nu}) \, +\, {\rm (regulator)},
\end{align}
where $g_{\mu\nu}$ and $K_{\mu\nu}$ are the metric and extrinsic curvature on a surface at constant $r$. The boundary metric $h_{\mu\nu} = \lim_{r\to\infty} g_{\mu\nu}$ is given by
\begin{align}
h_{\mu\nu} dx^\mu dx^\nu=- dt^2 + \fr{n} dx^i dx_i = -dt^2 + \fr{n} \gamma_{ab} d\theta^a d\theta^b.
\end{align}
where $(x^i,h_{ij}=\delta_{ij}/n)$ are the Cartesian and $(\theta^a,h_{ab}=\gamma_{ab}/n)$ are the lattice-adapted oblique coordinates with fixed period $\theta^a \sim \theta^a + 2\pi$, respectively. Note that all the lattice configuration is encoded in the oblique metric $\gamma_{ab}$. At the large $D$ limit, the quasi-local stress tensor in the Cartesian coordinates up to the leading order in $1/D$ is given by
\begin{align}
& T^{tt} = m,\quad T^{ti} = p^i-\partial^i m,\quad T^{ij} = (- m +\partial_t m + \partial_k p^k)\delta^{ij}
 - 2 \partial^{(i}p^{j)} + \frac{p^i p^j}{m},
\end{align}
where the indices $i,j$ are raised by $\delta_{ij}$. The components are normalized so that they remain finite at large $D$
\begin{align}
 \mathbf{T}_{\mu\nu} =\frac{(n+1)\Omega_{n+1}}{16\pi G} T_{\mu\nu}.
\end{align}
The ADM mass for a unit cell is given by
\begin{align}
 \textsc{Mass} =\int_{\rm cell} \mathbf{T}^{tt} \frac{d^px}{n^{p/2}}=
  \frac{(n+1)\Omega_{n+1}}{16 \pi G n^{p/2}} {\cal M}
\end{align}
where ${\cal M}$ is the normalized mass~(\ref{eq:def-normalizedmass}).
On the other hand, the variation of the ADM mass with respect to the spatial boundary metric $h_{ij}$ is given by~\cite{Brown:1992br,Donos:2013cka}
\begin{align}
 \frac{\delta }{\delta h_{ij}}\textsc{Mass}  =-  \fr{2} \int_{\rm cell}\mathbf{T}^{ij}\frac{d^px}{n^{p/2}}=\frac{(n+1)\Omega_{n+1}}{16 \pi G n^{p/2}} {\cal T}^{ij},
\end{align}
where ${\cal T}^{ij}$ is the normalized tension~(\ref{eq:def-tensionij}).
Thus, we obtain the relation between the normalized quantities
\begin{align}
 {\cal T}^{ij} = \frac{\delta \cM}{\delta h_{ij}}.
\end{align}
The brane tension in the oblique coordinates gives the mass variation with respect to the lattice parameter. In the $(u,v)$-coordinates~(\ref{eqn:uv_coords}), we have
\begin{align}
\delta_{k,\gamma} {\cal M} =\left.\fr{n} {\cal T}^{ab} \frac{\delta \gamma_{ab}}{\delta k}\right|_{\gamma} \delta k
+\left.\fr{n} {\cal T}^{ab} \frac{\delta \gamma_{ab}}{\delta \gamma}\right|_k\delta \gamma
\end{align}
where
\begin{align}
 \gamma_{ab} d\theta^a d\theta^b =\frac{(du^2+2\gamma dudv+dv^2)}{ k^2(1-\gamma^2)}.
\end{align}
The bulk tension~(\ref{eq:def-bulktension}) is related to the conformal change in the metric
\begin{align}
\left. \delta {\cal M} \right|_\gamma={\cal T} \frac{\delta (k^{-2})}{k^{-2}}.
\end{align}

\section{Black stripes}
In the two dimensional lattice setup, we also have branches with non-uniformity only in one direction, which we call {\it black stripes}. These branches play an important role in the non-equiangular lattice phases. 
The relevant branches of black stripes bifurcate from the following perturbations in the oblique coordinates $(u,v)$ in eq.~(\ref{eqn:uv_coords})\footnote{We do not consider the $(-)$-stripes branching off from $\cos(u-v)$ mode with $k=1/\sqrt{2(1+\gamma)}$, since they always have lower entropy for $0\leq \gamma\leq 1$. For $-1\leq \gamma\leq 0$, the $(+)$ and $(-)$-branches switch their roles.}
\begin{align}
\delta \cR\ \propto \ \cos u,\ \cos v,\ \cos(u+v),
\end{align}
where the first two modes have $k=1$ and the last $k=1/\sqrt{2(1-\gamma)}$.
The branches from $\cos u$ or $\cos v$, which we call $(0)$-stripes,
take the form $\cR(u,v)=F(u) \,{\rm or}\, F(v)$. Plugging this into eq.~(\ref{eqn:soap_bubble}),  we obtain the dimensionless version of the black string effective equation~\cite{Emparan:2015hwa,Suzuki:2015axa,Emparan:2018bmi}
\begin{align}
 F''(u)+\frac{1}{2}F'(u)^2+\fr{k^2} F(u)=0. \label{eq:nubs}
\end{align}
The other branch, the $(+)$-stripes, takes the form $\cR(u,v) = \tilde{F}(u+v)$.
$\tilde{F}(u)$ follows the same equation with $k\to k\sqrt{2(1-\gamma)}$.
Now, let us assume $F_k(u)$ as the non-uniform solution of eq.~(\ref{eq:nubs}). We define the following integrals
\begin{align}
 \mu_{\rm NUBS}(k) := \fr{2\pi} \int_0^{2\pi} e^{F_k(u)} du,\quad
  \tau_{\rm NUBS}(k) := \fr{2\pi} \int_0^{2\pi} F_k(u) e^{F_k(u)} du.
\end{align}
Then, the normalized entropy and relative tension of the black stripes are given by
\begin{align}
{\cal S}_{1,{\rm stripe}} = {\cal S}_{1,{\rm UBB}}-\log \left(\mu_{\rm NUBS}(\tilde{k})\right),\quad \tau_{\rm stripe} = 1 - \frac{\tau(\tilde{k})}{\mu(\tilde{k})}
\end{align}
where $ {\cal S}_{1,{\rm UBB}} = -\log ((2\pi)^2/k^{2}\sqrt{1-\gamma^2})$ and
\begin{align}
\tilde{k} = \left\{\begin{array}{cc}k&{\rm for}\ (0)-{\rm stripes}\\k\sqrt{2(1-\gamma)}& {\rm for}\ (+)-{\rm stripes}
 \end{array}\right..
\end{align}

\bibliographystyle{JHEP}
\bibliography{LargeDLatticesReferences}

\end{document}